\documentclass[runningheads]{llncs}
\usepackage[utf8]{inputenc}
\usepackage{booktabs}
\usepackage{graphicx}
\graphicspath{ {./figures/} }
\usepackage[graphicx]{realboxes}
\usepackage{array}
\usepackage{multirow}
\usepackage{longtable}
\usepackage{hyperref}
\usepackage{xcolor}

\hypersetup{backref=true,       
    pagebackref=true,               
    hyperindex=true,                
    colorlinks=true,                
    breaklinks=true,                
    urlcolor= black,                
    linkcolor= blue,                
    bookmarks=true,                 
    bookmarksopen=false,
    filecolor=black,
    citecolor=blue,
    linkbordercolor=blue
}

\begin{document}

\title{Blockchain Application Development\\Using Model-Driven Engineering and\\Low-Code Platforms: A Survey}

\titlerunning{Blockchain Application Development}

\author{Simon Curty\inst{1}\orcidID{0000-0002-2868-9001} \and
Felix Härer\inst{1}\orcidID{0000-0002-2768-2342} \and
Hans-Georg Fill\inst{1}\orcidID{0000-0001-5076-5341}}
\authorrunning{S. Curty et al.}
%
\institute{University of Fribourg, Digitalization and Information Systems Group, Fribourg, CH
\email{firstname.lastname@unifr.ch}\\
\url{https://www.unifr.ch/inf/digits/}}
\maketitle              
\begin{abstract}
The creation of blockchain-based software applications requires today considerable technical knowledge, particularly in software design and programming. This is regarded as a major barrier in adopting this technology in business and making it accessible to a wider audience. As a solution, no-code and low-code approaches have been proposed that require only little or no programming knowledge for creating full-fledged software applications. In this paper we review academic approaches from the discipline of model-driven engineering as well as industrial no-code and low-code development platforms for blockchains. We further present a case study for an integrated no-code blockchain environment for demonstrating the state-of-the-art in this area. Based on the gained insights we derive requirements for the future development of no-code and low-code approaches that are dedicated to the field of blockchains.

\end{abstract}
\section{Introduction}
\label{sec:introduction}

With the further maturing of blockchain technologies and the soon expected transition to more energy-efficient and faster protocols with higher transaction volumes~\cite{kim_2021,Fairley2019,Nguyen2019}, a more widespread adoption of these technologies seems within reach. However, one considerable barrier limiting the adoption is the technical and organizational complexity that users are confronted with when creating blockchain-based applications~\cite{HolotiukM18}. This complexity originates on the one hand from the underlying technical foundations, which build on distributed and decentralized systems, cryptography, and algorithmic processing~\cite{antonopoulos2018}. Blockchains such as Ethereum combine these properties for storing transactions in an append-only data structure, where each new block has a cryptographically verifiable link to its predecessor. Thus, users are part of a decentralized network that minimizes the degree of trust required towards other participants who continuously validate the links of the blockchain. In addition, organizational barriers such as the involvement of new regulatory requirements, the development of new skills and competencies, and the availability of financial and human resources may prevent adoption in practice~\cite{Clohessy2019}. \newline
From the perspective of software engineering, the lack of specialists for programming may today be partly compensated with so-called \emph{low-code} platforms \cite{fill2021,Tisi2019,bockLowCodePlatform2021}. These development platforms are typically available as cloud services with visual, diagrammatic interfaces and declarative languages. In our view, they constitute the next step in the industry adoption of academic model-driven engineering (MDE) approaches and its predecessors where models are regarded as primary development artifacts for software engineering~\cite{whittle2013,brambillaModelDriven2017,diruscio_low-code_2022}.
While \textit{low-code} approaches allow a user to produce results without having to understand source code and there may be an underlying model integrated with features of the platform~\cite{bockLowCodePlatform2021}, the model may not conform to an explicit formalization~\cite{diruscio_low-code_2022}. Further, we consider so-called \textit{no-code} approaches as a subset of low-code approaches that operate at an abstraction level above code, not showing code to the user at all. Today, a large number of such platforms and tools are available that either support the development of complete software applications or focus on providing specific functionality, e.g.\ for entering data in a form and saving it to a database~\cite{sahay2020}. \newline
For easing the creation of blockchain-based applications it seems obvious to revert to MDE and low-code approaches. These carry the potential to abstract from the technical complexity and enable users to focus on usage scenarios and the organizational embedding. In the following we investigate academic and industrial approaches for realizing blockchain applications using these methods. We will do this along the following three research questions.
\emph{RQ1:} Which academic MDE approaches exist for the development of blockchain-based applications?, \emph{RQ2:} Which low-code and no-code platforms permit the realization of blockchain-based applications?, \emph{RQ3:} What are requirements for future blockchain development platforms that are informed by MDE, no-code and low-code? \newline
In particular, we will regard approaches that are already available for creating blockchain-based software applications or offer interfaces to other platforms enabling this. This will permit to describe the state-of-the-art in this area and derive requirements for the development of future approaches. The remainder of the paper is structured as follows. Section 2 will outline related work in the form of previous studies and lead over to our research methodology in Section 3. Subsequently, we will present in Section 4 our review of academic MDE approaches and in Section 5 the review of no-code and low-code development platforms used in industry. Section 6 presents a blockchain use case using state-of-the-art low-code platforms, resulting in the discussion of requirements in Section 7.

\section{Related Studies}
Developing blockchain-based applications requires a high level of expertise and understanding of the underlying technologies. Blockchain-based applications are empowered by smart contracts, i.e.\ programs executed on the blockchain. These smart contracts often involve financial transactions or deal with issues related to trust. As such, their correctness is of utmost importance. Due to the immutable nature of blockchains, mistakes in smart contract implementations are difficult to rectify.
This can be eased through different visual languages for smart contracts, which have been reviewed and compared in~\cite{harerComparisonApproachesVisualizing2019}. While visual programming languages aim to reduce complexity and improve accessibility for the programmer, they do not correspond in general to low-code development approaches, which may involve visual programming but also deal with the generation and life-cycle management of software artifacts. Approaches and tools for the analysis and development of smart contracts have been reviewed in~\cite{HU2021100179,VACCA2021110891}. While both studies discuss issues related to software engineering, such as code analysis and testing, model-driven or low-code techniques to develop blockchain-based software are not regarded. 

The study by Ait Hsain et al.~\cite{AITHSAIN2021785} focuses on MDE for Ethereum smart contracts, however the review process is not elaborated. Sánchez-Gómez et al.~\cite{9186040} review model-based testing and development approaches. Since the publication of their study, newer approaches have emerged. A more recent review of MDE methods was conducted by Levasseur et al.~\cite{levasseur_survey_2021}. In comparison to their work, we applied a broader search methodology and identified more approaches.
None of these studies consider industrial approaches such as platforms and focus predominantly on smart contracts.

In summary, while numerous studies on issues regarding smart contract development have been conducted, to the best of our knowledge, a comprehensive review of the state-of-the-art of MDE and low-code/no-code approaches from both academia and industry in this field is missing so far.

\section{Research Methodology}

For answering the three research questions we will employ the following research methodology. At first we  review existing academic MDE approaches for blockchain applications in the form of a structured literature review (SLR). Thereby we follow the guidelines by Webster and Watson~\cite{WebsterW02} and vom Brocke et al.~\cite{BrockeSRNPC15}. The initial corpus of the SLR was generated by searching all keyword combinations from two groups, where group one included '\textit{blockchain, distributed ledger, smart contract}' and group two '\textit{enterprise model, conceptual model, business model, model-driven, no-code, low-code}'. These keywords were selected based on the domain understanding of the authors. We expected the relevant concepts to be dispersed, thus we chose a broad set of keywords.

For discovering relevant industrial approaches, we reverted to expert knowledge from industry in the field of low-code development combined with our own searches. On this bases, we conducted (1) a survey of available platforms towards suitability for blockchain application development and (2) the implementation of a blockchain use case as an evaluation. This exploratory research approach is directed towards discovering requirements for future platforms that combine blockchain application development with the state-of-the-art from academia and industry.

\section{Academic MDE Approaches}

In the following subsections, we review approaches of the academic discipline \textit{model-driven engineering} in regard to development solutions for blockchains.

Model-driven engineering introduces models as primary artifact to the software development process in order to address numerous challenges of software engineering~\cite{brambillaModelDriven2017,Schmidt2006}: First, the common understanding of software artifacts can be facilitated by domain-specific models, as such models are easier to interpret for humans than code. Second, model-based reasoning allows the verification of software, e.g., to determine the fulfillment of security properties. And third, well-defined models allow developers to create software artifacts in an automated fashion, which are correct-by-construction, with no or reduced coding effort.
To identify existing MDE approaches that target specifically the development of blockchain applications, we conducted a systematic literature review as elaborated in the following.

\subsection{Review Process}
\label{subsec:ac_review_process}
\begin{figure} [htbp]
    \centering
    \includegraphics[width=0.75\linewidth]{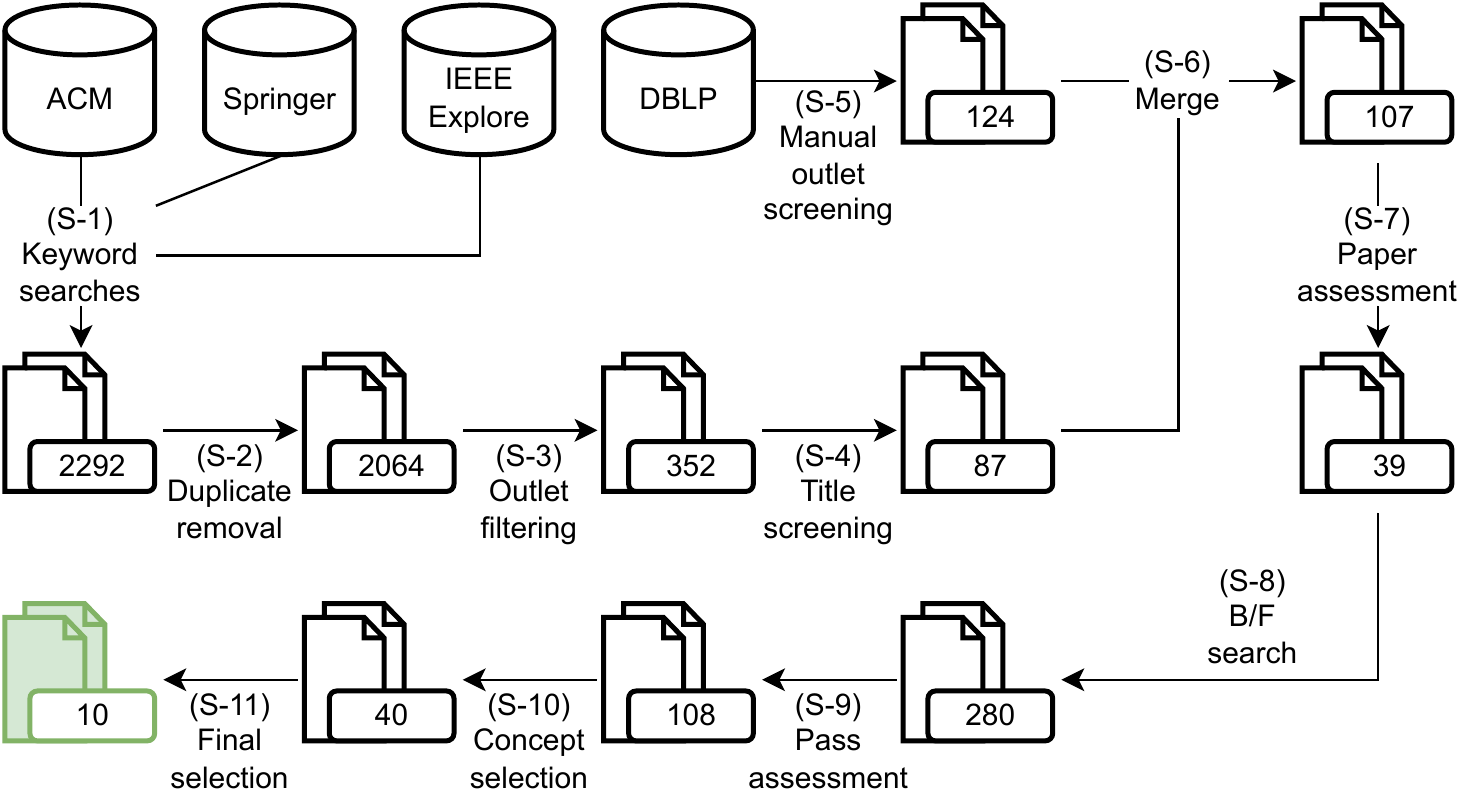}
    \caption{Academic Literature Review Process}
    \label{fig:ac_review_process}
\end{figure}

The systematic review process as shown in Figure~\ref{fig:ac_review_process} follows the guidelines by~\cite{WebsterW02} and~\cite{BrockeSRNPC15}. 
To obtain an initial corpus of publications, we performed keyword searches in step (S-1) on ACM, Springer, and IEEE Explore with the search strings shown in Table~\ref{tab:search_strings}. From the resulting corpus, duplicates were removed in step (S-2). Due to the large number of documents, we filtered the publications by outlets in step (S-3) that typically publish papers in software engineering, model-driven engineering or information systems.
Before the fulltext analysis, the reduced corpus was then screened by titles in step (S-4). 

As basis for this fourth step we formulated keyword criteria, whereby the title should contain one of \textit{"conceptual", "model", "process", "execution", "process", "architecture", "framework", "design", "development", "pattern", "use case", "supply chain", "database", "storage", "verification", "generation", "language"}, and mention a blockchain-related word, such as \textit{"distributed", "chain", "contract"}. Additionally, we analyzed all titles  to capture promising publications. In parallel, we screened in step (S-5) the table of contents of selected outlets in software engineering and related disciplines by applying the same process as in (S-4). That is, we screened all proceedings, workshops, issues, etc., published in one of the following outlets from 2015 to Nov. 2021: \textit{BCCA, BMSD, BPMDS/EMMSAD, BRAINS, COINS, CSIMQ, CVCBT, DAPPCON, DK, EMISA, ER, ICBC, ICBCT, IEEE Blockchain, IEEE ICBC, IJISMD, MoDELS, PoEM} and \textit{SoSyM}. These two sets of publications were then merged and duplicates removed (S-6). 

\begin{table}[]
\scriptsize
\centering
\begin{tabular}{p{0.9\linewidth}|r}
\toprule
Search string & Results \\
\midrule
("blockchain" OR "distributed ledger" OR "smart contract" OR "smart-contract") AND ("All "business model" OR "business modeling") AND (year\textgreater 2014) & 1625 \\
("blockchain" OR "distributed ledger" OR "smart contract" OR "smart-contract") AND ("All "enterprise model" OR "enterprise modeling") AND (year\textgreater 2014) & 40\\
("blockchain" OR "distributed ledger" OR "smart contract" OR "smart-contract") AND ("All "conceptual model" OR "conceptual modeling") AND (year\textgreater 2014) & 370 \\
("blockchain" OR "distributed ledger" OR "smart contract" OR "smart-contract") AND ("All "model driven" OR "model-driven") AND (year\textgreater 2014) & 181 \\
("blockchain" OR "distributed ledger" OR "smart contract" OR "smart-contract") AND ("All "no code" OR "no-code" OR "low code" OR "low-code") AND (year\textgreater 2014) & 76 \\
\bottomrule
\end{tabular}
\caption{Simplified search strings used and results found on ACM, IEEE Explore, and Springer. The concrete syntax of search strings varies for each search portal.}
\label{tab:search_strings}
\vspace{-8mm}
\end{table}

In the next step, the publications were assessed by at least reading the abstract and reviewing tables and images (S-7), considering the inclusion criteria that (i) the publication should be directly related to distributed ledger technologies, and (ii) creates, discusses, or presents a modeling approach. Publications using models to only illustrate software, systems, or a use case, e.g., by means of a standard UML use case diagram, were excluded. For the remaining publications, we then performed a recursive backward-forward search, as proposed in~\cite{WatsonW20} (S-8): references and citations were screened, seemingly relevant publications added to the set, and subsequently assessed as in step (S-7). For all relevant new additions, a backward-forward search was again performed. \newline
Eventually, no new relevant publications could be found and the backward-forward search was concluded. Of all thus collected publications, 108 fulfilled the assessment criteria (S-9). We further filtered by contained concepts in step (S-10), i.e., (i) the approach has MDE characteristics, (ii) it must be tool-assisted, and (iii) include generation of code, application artifacts, or some executable specifications. The motivation for choosing these criteria is founded in the commonalities of low-code/no-code and MDE, as elaborated in Section~\ref{sec:introduction}. Finally, we selected 10 approaches we consider representative for the full spectrum of academic approaches.

\subsection{Results}

In Table \ref{tab:academia-results}, the final selection of academic approaches from (S-11) is shown. We further evaluated the approaches regarding the required user expertise - see column \emph{Expertise}. Approaches where a user must not write any code and only basic understanding of blockchain concepts is required, we consider suitable for \textit{non-technical} users. In contrast, approaches that require understanding of advanced concepts or chain-specific features, e.g. gas costs in Ethereum, we consider suitable for \textit{non-programmers}. Finally, if the user has to write any code, the approach is only suitable for \textit{programmers}.

For the comparison of the academic approaches, we classified them in addition using three layers, which are based on the traditional layers of ArchiMate\footnote{See \url{https://www.opengroup.org/archimate-forum/archimate-overview}}. This choice is motivated by a previous application of ArchiMate in the context of Blockchain use cases~\cite{Curty2021TowardsTC}. Approaches on the \textbf{(i) business layer} integrate modeling of business concepts, such as use cases from a top-down perspective. The \textbf{(ii) application layer} includes approaches which integrate life-cycle and deployment management, or integration facilities. Finally, on the \textbf{(iii) technology layer}, we consider approaches whose scope is limited to the generation of smart contract code from models.

\bgroup
\setlength{\tabcolsep}{0.3em}
\begin{table}[]
\scriptsize
\centering
\begin{tabular}{@{}lllp{1.9cm}l>{\raggedright\arraybackslash}p{2.5cm}rr@{}}
\toprule
Ref.     & Name                     & BP    & Modeling language                 & Layer         & Impl. platform                & Expertise  & OS  \\ \midrule
{\cite{archi2hc}} & Archi2HC        & H     & ArchiMate                         & Business      & Archi                         & •••   & - \\
{\cite{caterpillar}} & Caterpillar  & E     & BPMN                              & Application   & custom (Node.js, bpmn-js)     & •oo   & + \\
{\cite{chainops}} & ChainOps        & E     & domain-specific                   & Application   & AstraKode Blockchain Modeler  & •oo   & - \\
{\cite{dascontract}} & Das Contract & E,C   & DEMO, BPMN, Blockly               & Technology    & custom (.Net, Node.js)        & ••o   & o \\
{\cite{icontractbot}} & iContractBot& MC    & domain-specific (iContractML)     & Technology    & Xatkit                        & •oo   & o \\
{\cite{icontractml}} & iContractML  & MC    & domain-specific (iContractML)     & Technology    & Obeo Designer (Eclipse Sirius)& •oo   & o \\
{\cite{latte}} & LATTE              & E     & domain-specific                   & Technology    & custom (Electron)             & ••o   & + \\
{\cite{lemma}} & LEMMA              & E     & domain-specific (LEMMA)           & Application   & LEMMA                         & •••   & o \\
{\cite{uml2go}} & UML2Go            & H     & UML                               & Technology    & Obeo Acceleo (Eclipse)        & ••o   & - \\
{\cite{verisolid}} & VeriSolid      & E     & domain-specific (state machine)   & Technology    & WebGME                        & ••o   & + \\
\bottomrule
\multicolumn{8}{l}{Name: Short name of the approach. If none was given by the authors we assigned one.} \\
\multicolumn{8}{l}{BP: Blockchain platform, E: Ethereum, C: Cardano, H: Hyperledger Fabric, MC: Multi-chain}\\
\multicolumn{8}{l}{Exp.: Required expertise, •: non-technical, ••: non-programmer, •••: programmer}\\
\multicolumn{8}{l}{OS: open source, +: available, o: no license specified, -: not available}
\end{tabular}
\caption{Selected academic, model-driven approaches for blockchain application development that apply code generation.}
\label{tab:academia-results}
\vspace{-8mm}
\end{table}
\egroup

In the entire corpus of publications, we could identify only one approach which clearly lies on the \textbf{business layer (i)} and simultaneously permits the generation of code artifacts. Babkin et al.~\cite{archi2hc} propose a mapping between ArchiMate concepts and Hyperledger Composer constructs. From an ArchiMate model, a project artifact for Hyperledger Fabric is generated. However, a programmer must implement the business logic manually. The Das Contract approach~\cite{dascontract} applies modeling languages of DEMO to design and generate smart contracts. While DEMO is traditionally used to model organizations, this is however not part of of this approach.\newline 
On the \textbf{application layer (ii)} lie approaches and tools offering integration and management capabilities for the generated artifacts. Caterpillar~\cite{caterpillar} is a process execution system, in which processes are modeled as BPMN in a web-based visual editor. The models may be translated to Solidity code or into an intermediate representation to be executed by an on-chain execution engine. Furthermore, the tool offers a model repository and monitoring of processes. In the ChainOps~\cite{chainops} framework, smart contracts are composed visually from pre-defined templates, subsequently validated against domain-specific constraints and policies. Models then are sent to a REST service to be translated and deployed. The vision of ChainOps is to offer a complete and integrated Dapp life-cycle solution for the OntoChain\footnote{\url{https://ontochain.ngi.eu/}} ecosystem. The work of Trebbau et al.~\cite{lemma} is an extension of LEMMA, a modeling framework for microservices. Using the modeling languages of LEMMA, code artifacts for the connection to chain networks and smart contract interaction may be generated. The focus of this approach is the model-based integration of on-chain components.\newline
Most identified approaches focus on the generation of smart contract code without offering additional life-cycle capabilities, and are thus assigned to the \textbf{technology layer (iii)}. Suitable for non-technical users are approaches which abstract blockchain and platform-specific concepts. The modeling language iContractML~\cite{icontractml} has a visual notation with few elements for the specification of the structure of smart contracts. Models are translated to DAML, which is compatible with various chains. Based on this language, iContractBot~\cite{icontractbot} allows the user to specify models conversationally. Another approach targeting multiple chains is the aforementioned Das Contract, in which the behavior of a contract is specified in Blockly. Since Blockly contains coding concepts, we do not consider it suitable for non-technical users.
Approaches specifically for Ethereum are LATTE~\cite{latte} and VersiSolid~\cite{verisolid}. The former relies on a combination of form-based definition of the structure of Solidity contracts and their implementation, defined visually in a notation similar to flow-charts. In the latter, Solidity contracts are modeled as state machines in visual fashion. This approach focuses on the formal verification of the generated contract. Another platform-specific approach is UML2Go~\cite{uml2go} for Hyperledger Fabric. Contracts are modeled as UML class and sequence diagrams and then translated to Go chaincode using model transformation. \newline
The results show that academic approaches predominantly focus on the platform-specific generation of smart contract code, while holistic solutions are sparse.

\section{Industrial Low-Code and No-Code Approaches}
For practitioners, an increasing number of low-code and no-code solutions is available.
In an informal compilation by Invernizzi and Tossell\footnote{https://pinver.medium.com/decoding-the-no-code-low-code-startup-universe-and-its-players-4b5e0221d58b}, solutions from 145 companies were found, such as website and app builders, e-commerce services, and data dashboards. The identified solutions differ substantially in the scope and applications they target.
With the aim of assessing the scope and applicability of industrial approaches towards blockchain applications, we conducted a review of state-of-the-art solutions. The data sources for this review are (DS-1): the compilation by Invernizzi and Tossell, (DS-2): practical approaches from prior research \cite{harerComparisonApproachesVisualizing2019}, and (DS-3): additional research on blockchain-specific no-code and low-code solutions available on the web. 

\subsection{Review Process}

We applied a three-step process, consisting of an initial filtering step (S-1), the evaluation of scope and applicability for blockchains in step (S-2), and  the classification of solutions applicable to blockchains (S-3). Initially, 169 solutions were identified. In (S-1), we manually retrieved descriptions from the vendor websites in addition to information provided by (DS-1), followed by filtering out duplicate entries, those that could not be reached on the web, or did not provide sufficient information on their websites (e.g.\ closed beta software). The remaining 150 solutions were evaluated in (S-2) regarding their scope of blockchain integration. Finally, 40 solutions were identified as applicable for blockchains.

\subsection{Results}
For discussing available platforms and their blockchain integration, we distinguish between \emph{1st degree} and \emph{2nd degree} integration. A platform supports 1st degree integration if it interacts directly with blockchains through its software or services. 2nd degree integration is supported if an external service could be integrated that offers 1st degree integration. The criteria for the selected platforms (S-3) listed in Table \ref{tab:table1} are that they (a) offer blockchain integration of 1st or 2nd degree and (b) were considered a low-code or no-code approach.

\begin{table}[ht]
\centering
\scalebox{0.865}{
\begin{tabular}{@{}rrllccclrrllccc@{}}
\cmidrule(r){1-8} \cmidrule(l){8-15}
&
Cat.&
Name&
Website&
d\textsubscript{1}&
d\textsubscript{2}&
 s\rule{0.8ex}{0pt}&
 &
 &
Cat.&
Name&
Website&
d\textsubscript{1}&
d\textsubscript{2}&
s \\ \cmidrule(r){1-8} \cmidrule(l){8-15} 
1  & AM & Adalo       &\href{http://www.adalo.com}{adalo\scriptsize{.com}}& - & + & o & & 21 & SC   & DAML         &\href{http://daml.com}{daml\scriptsize{.com}}& +   & - & +\\
2  & AM & BuildFire   &\href{http://buildfire.com}{buildfire\scriptsize{.com}}& - & + & o & & 22 & SC   & Simba Chain  &\href{http://simbachain.com}{simbachain\scriptsize{.com}}& +   & - & - \\
3  & AM   & Glide        &\href{http://www.glideapps.com}{glideapps\scriptsize{.com}}& +   & +   & - & & 23 & SC & Dappbuilder &\href{http://dappbuilder.io}{dappbuilder\scriptsize{.io}}& + & - & +\\
4  & AM   & Axonator     &\href{http://axonator.com}{axonator\scriptsize{.com}}& - & +   & - & & 24 & SP   & Airtable     &\href{http://airtable.com}{airtable\scriptsize{.com}}& - & +   & - \\
5  & AW   & Outsystems   &\href{http://outsystems.com}{outsystems\scriptsize{.com}}& +   & - & - & & 25 & WA   & n8n          &\href{http://n8n.io}{n8n\scriptsize{.io}}& +   & - & +\\
6  & AW   & Builder.ai   &\href{http://builder.ai}{builder\scriptsize{.ai}}& +   & - & - & & 26 & WA & Zapier      &\href{http://zapier.com}{zapier\scriptsize{.com}}& + & + & o \\
7  & AW   & Bubble       &\href{http://bubble.io}{bubble\scriptsize{.io}}& +   & +   & - & & 27 & WA   & Integromat   &\href{http://www.integromat.com}{integromat\scriptsize{.com}}& +   & +   & - \\
8  & AW   & Landbot      &\href{http://landbot.io}{landbot\scriptsize{.io}}& - & +   & - & & 28 & WA   & Process Str. &\href{http://www.process.st}{process\scriptsize{.st}}& - & +   & - \\
9  & AW & Draftbit    &\href{http://draftbit.com}{draftbit\scriptsize{.com}}& - & + & o & & 29 & WA   & IFTTT        &\href{http://ifttt.com}{ifttt\scriptsize{.com}}& +   & +   & - \\
10 & D    & Parabola     &\href{http://parabola.io}{parabola\scriptsize{.io}}& - & +   & - & & 30 & WA   & NodeRed      &\href{http://nodered.org}{nodered\scriptsize{.org}}& +   & +   & +\\
11 & D    & Gyana        &\href{http://gyana.com}{gyana\scriptsize{.com}}& - & +   & o & & 31 & WA   & Aurachain    &\href{http://www.aurachain.ch}{aurachain\scriptsize{.ch}}& +   & - & - \\
12 & D    & Obviously AI &\href{http://www.obviously.ai}{obviously\scriptsize{.ai}}& - & +   & - & & 32 & WB   & Webflow      &\href{http://webflow.com}{webflow\scriptsize{.com}}& +   & +   & - \\
13 & D    & Levity       &\href{http://levity.ai}{levity\scriptsize{.ai}}& - & +   & - & & 33 & WB   & Unstack      &\href{http://www.unstack.com}{unstack\scriptsize{.com}}& - & +   & - \\
14 & F  & Arengu      &\href{http://arengu.com}{arengu\scriptsize{.com}}& - & + & o & & 34 & WB   & Squarespace  &\href{http://www.squarespace.com}{squarespace\scriptsize{.com}}& - & +   & - \\
15 & F    & Formstack    &\href{http://formstack.com}{formstack\scriptsize{.com}}& - & +   & - & & 35 & WB   & Linktree     &\href{http://linktr.ee}{linktr\scriptsize{.ee}}& - & +   & - \\
16 & F    & Tally        &\href{http://tally.so}{tally\scriptsize{.so}}& - & +   & - & & 36 & WB   & Pory         &\href{http://pory.io}{pory\scriptsize{.io}}& - & +   & - \\
17 & IN   & Budibase     &\href{http://budibase.com}{budibase\scriptsize{.com}}& - & +   & - & & 37 & WB   & Softr        &\href{http://www.softr.io}{softr\scriptsize{.io}}& - & +   & - \\
18 & IN   & Flowdash     &\href{http://flowdash.com}{flowdash\scriptsize{.com}}& - & +   & - & & 38 & WB & Xooa        &\href{http://xooa.com}{xooa\scriptsize{.com}}& + & - & o \\
19 & IN & Jet Admin   &\href{http://jetadmin.io}{jetadmin\scriptsize{.io}}& - & + & o & & 39 & WB & ICME        &\href{https://www.icme.io}{icme\scriptsize{.io}}& + & - & - \\
20 & IN & Windward    &\href{http://www.windwardstudios.com}{windwardstudios\scriptsize{.com}}& - & + & - & & 40 & WB   & Atra         &\href{http://atra.io}{atra\scriptsize{.io}}& +   & - & o \\ \cmidrule(r){1-8} \cmidrule(l){8-15} 

\multicolumn{15}{l}{Cat.: Category, AM: app builder with mobile focus, AW: app builder with web focus, D: data}\\
\multicolumn{15}{l}{F: forms, IN: internal tools, SC: smart contracts, SP: Spreadsheets, WA: workflow automation}\\
\multicolumn{15}{l}{WB: website builders, d\textsubscript{1}: 1st degree integration, d\textsubscript{2}: 2nd degree integration, s: open source}\\
\multicolumn{15}{l}{+: applicable, o: partially applicable, -: not applicable}\\
\multicolumn{15}{l}{}\\
\end{tabular}
}

\caption{Low- and no-code approaches with 1st or 2nd degree blockchain integration.}
\label{tab:table1}
\vspace{-8mm}
\end{table}

\textbf{Categories with 1st degree integration}: 1st degree blockchain integration has been found in 17 solutions intended for building websites and apps, workflow automation, and smart contract development. 
Exemplary integration features in the \emph{app builder} category are the creation of decentralized apps (Dapps) and the integration of cryptocurrency-related data, e.g.\ price information. App builders such as Outsystems (5) and Bubble (7) support Dapps, where components of a mobile, desktop, or web app can send blockchain transactions and call smart contract functions, e.g.\ through the MetaMask browser extension. \newline
For \emph{website builders}, blockchain integration has only been found for integrating cryptocurrency-related data, with the exception of ICME (39). ICME is a website builder for creating websites on the Dfinity blockchain. The app and the resulting websites are hosted on Dfinity.\newline
\emph{Workflow automation} tools allow for the execution of user-defined workflows. A workflow is entered via a visual flow-based editor, showing the subsequent flow of steps for execution along with execution logic, or using dialogs or forms. Exemplary integration features are transactions and smart contract support for the Ethereum blockchain in Zapier (26) and Aurachain (31), and support for the Hyperledger Fabric blockchain in NodeRed (30) and Aurachain (31). Further services support crypto-currency data integrations. \newline
\emph{Smart contract development} is supported by integration features in Hyperledger Fabric, Hyperledger Sawtooth, Amazon QLDB, and others in DAML (21), a domain-specific language for textual descriptions of smart contracts. The textual language uses a syntax with natural language elements that can be interpreted and deployed for the supported platforms. Smart contract design based on templates and a visual editor is found for Ethereum, Hyperledger Fabric, and others in SimbaChain (22). The editor supports the creation of smart contracts by defining assets and transactions. Dappbuilder (23) offers smart contract creation from pre-defined templates for Ethereum, Polygon, and others. The approach limits applicability to standardized contracts, e.g.\ issuing tokens according to the Ethereum ERC-20 and similar token standards.

\textbf{Categories with 2nd degree integration}: 2nd degree blockchain integration has been found in 30 solutions intended for building websites, apps, or forms, for workflow automation, internal tools for companies, and for data processing and spreadsheets. 7 solutions also offer 1st degree blockchain integration. The integration features across the categories rely on another services providing a direct integration for blockchain applications. Among the no-code or low-code applications, it is typical to integrate other services in the fashion of a composition, for example, creating an application in an app builder with data provided by an external service. Blockchain integration features, due to this capability, rely on other services for blockchain integration. \newline
Notably, 28 of the 30 solutions integrate with Zapier (26), thereby offering support for \emph{interacting with Ethereum smart contracts and transactions}. These concern website and app builders such as Glide (3), which can embed dialogs for smart contracts and transactions in this way, in addition to integrating cryptocurrency data. Similarly, form builders allow defining input fields and the processing of submitted data through integrations. Arengu (14) is a typical example which also supports visualization with a flow-based editor.\newline
\emph{Workflow automation} tools offer the integration as part of the executable workflow definition. For example, a transaction may be sent after the workflow has been started by another action such as entering data in a spreadsheet. This is often accomplished by integrating AirTable (24). Internal tools include software tools for enterprises, automating typical enterprise resource tasks or operational tasks, e.g.\ using JetAdmin (19), or business processes as in Flowdash (18). Integrations in this context can be triggered similar to workflow automation tools.\newline
\emph{Data processing and spreadsheets} tools permit integrating data sources, thereby enabling for example the processing of newly appearing blockchain transactions, filtering for specified criteria, and calculations such as the aggregation of transferred amounts. Examples where this is possible are the spreadsheet tool AirTable (24) and data analytics tools such as Parabola (10).

The results show that the integration possibilities for the creation of websites or apps hinge on few services such as Zapier (26), predominantly found in the workflow automation category. Typical integration features consist of access to blockchain transactions or cryptocurrency data. Further integration possibilities with APIs on a technical level are very common, however, they were not considered no-code or low-code when using using Webhooks, Rest, other forms of HTTP requests or technical API descriptions. For the development of smart contracts, few no-code instances could be found in practice, with all of them requiring expert knowledge in blockchains.

\section{Use Case for Low-Code Blockchain Development}

For conducting a first evaluation of the state-of-the-art in realizing blockchain applications using low-code and no-code approaches (RQ2), we implemented a blockchain app with a smart contract in the area of supply chain tracking and tracing. For this purpose, we selected the Outsystems low-code platform, which targets developers, together with the SimbaChain platform as an exemplary no-code platform that is directed towards end-users.

The goal was to provide a trusted and up-to-date IT system shared by distributed supply chain participants.
In this domain, blockchain-based solutions promise information that is available as a trusted source in near-time or real-time among network participants~\cite{heloRealtimeSupplyChain2020,chenTransactionCostPerspective2022}. In particular, the tracking of goods in international shipments is a challenging area, involving the coordination of material flows from suppliers and manufacturers through container and sea freight companies to distributors. Additionally, products and materials need to be traced back to their source. Without a trusted IT infrastructure, shipments are mostly documented using paper documents, with point-to-point communication by e-mail, phone, and siloed IT systems, resulting in high transaction costs~\cite{zengAdoptionOpenPlatform2020,chenTransactionCostPerspective2022}. 

\begin{figure}[!htb]
  \centering
  \includegraphics[width=0.79\linewidth]{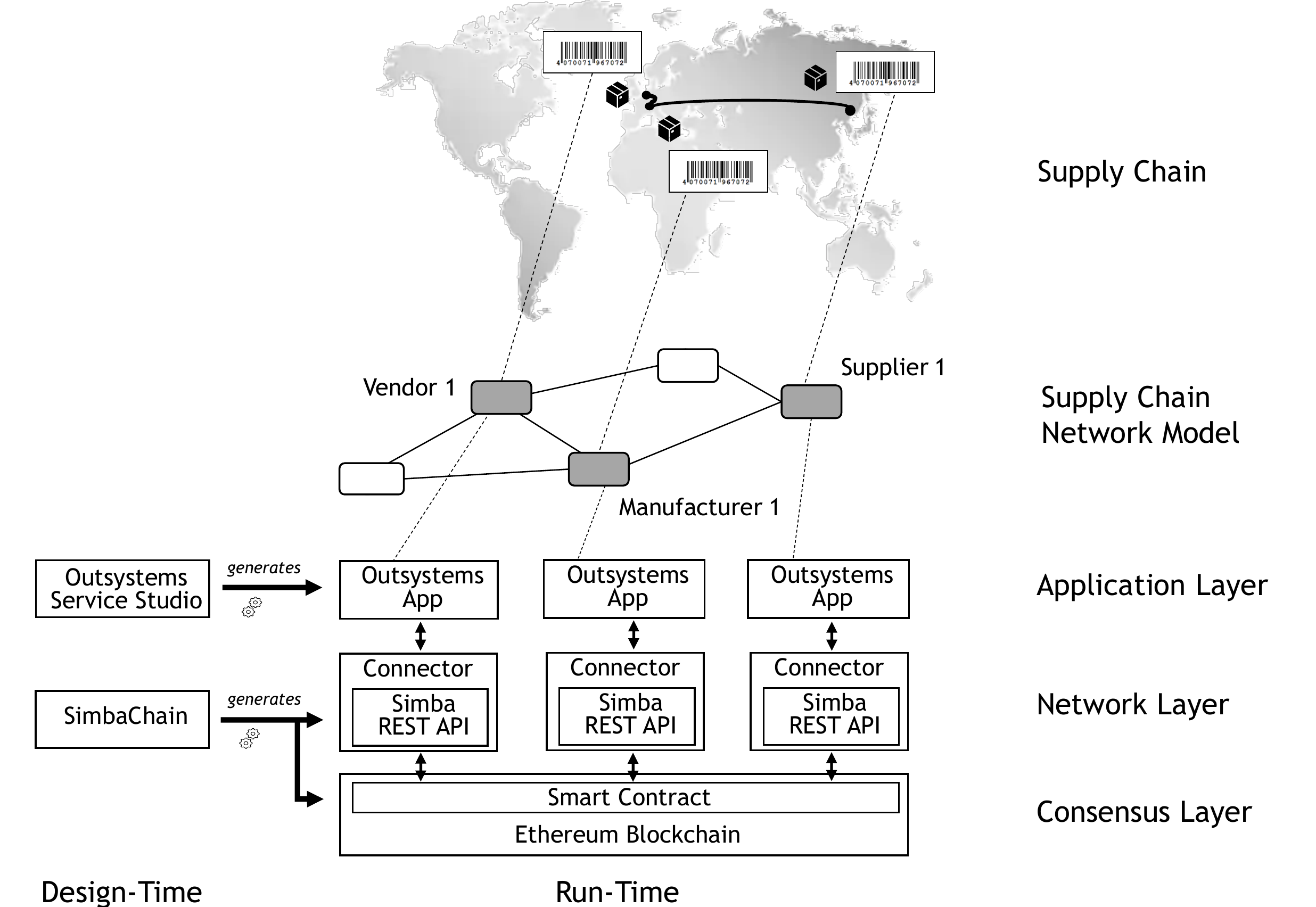} 
  \caption{Blockchain-based Architecture for Supply Chain Tracking and Tracing.}
  \label{fig:uc_architecture}
\end{figure}

\begin{figure}[!hb]
  \centering
  \includegraphics[width=0.79\linewidth]{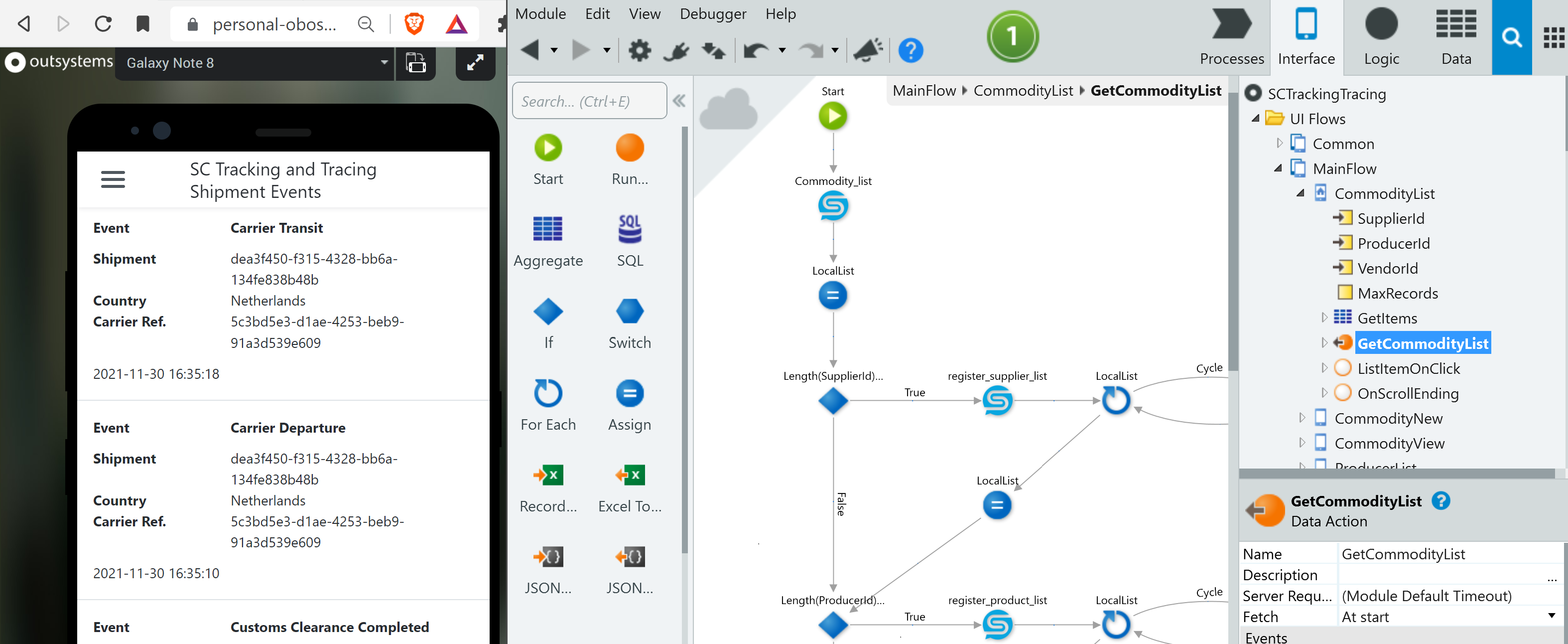} 
  \caption{Design of mobile app (left-hand side) in Outsystems Service Studio (right-hand side) using a flow-based editor for processing commodity data records of shipments.}
  \label{fig:uc_app}
\end{figure}

\begin{figure}[!htb]
  \centering
  \includegraphics[width=0.69\linewidth]{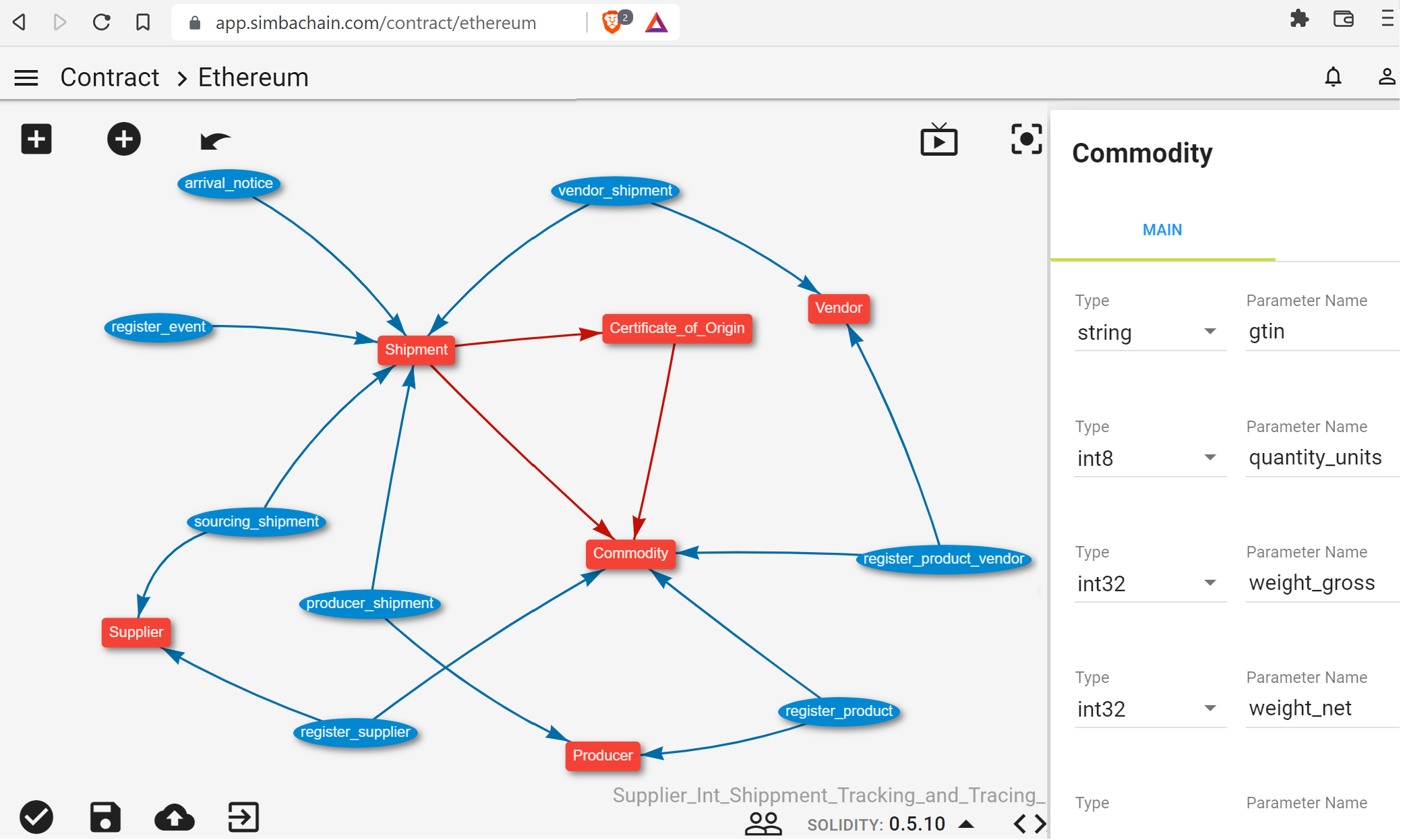} 
  \caption{Smart contract design in SimbaChain using a visual editor for a data model of transactions (blue) and assets (red).}
  \label{fig:uc_smart_contract}
\end{figure}

Figure~\ref{fig:uc_architecture} shows the implemented three-layer architecture with an exemplary network. Using Outsystems studio, an app was designed for registering suppliers, manufacturers, and vendors together with freight forwarding companies, commodities, and shipments. On the application layer, Supplier 1 might scan a shipment through a Global Trade Item Number (GTIN) with a smartphone camera and submit related IDs and attributes. For Manufacturer 1 and Vendor 1, this data becomes available and is updated with shipment events by freight providers and forwarders. Figure~\ref{fig:uc_app} shows this data in the app during development. The Ethereum blockchain is integrated for establishing a consistent view of data on the network and consensus layers of the architecture. In Outsystems, a REST API hosted by Simba relays requests to the Ethereum smart contract. Smart contracts and APIs are generated together through SimbaChain by specifying the model in Figure~\ref{fig:uc_smart_contract}. 

\section{Discussion and Requirements for Future Developments}

The review of academic approaches for model-driven engineering for blockchain applications has shown that most approaches focus on the technical level and not all of them support code generation. Rather, many approaches target the formal verification of smart contracts and only some approaches provide working prototypes. Approaches that integrate the business, application, and technical layer have not yet been proposed prominently in the literature. These would however bring benefits in terms of a holistic view on blockchain application design and should be investigated in the future.

The reviewed no-code and low-code approaches as used in industry showed the high maturity of these platforms. This concerned in particular the high usability, the availability of a broad range of interfaces for cloud-based and blockchain integrations and the possibility of cross-platform development. On the downside, it is hard to trace errors and debug applications on some low-code platforms as implementation details are hidden. Although some platforms offer the inspection of the generated code, this requires again technical know-how.

The practical use case permitted further insights.  Regarding the blockchain implementation through SimbaChain, a major architectural limitation is the generation of APIs used as relay when accessing the smart contract. Additional validation of the blockchain is required for assessing the consistency of data. 
From a user perspective, the SimbaChain platform requires only high-level knowledge of data types in addition to the visual entity concept documented in the platform. While SimbaChain might thus be considered a \textit{no-code} approach, it is limited to the presented operations and its resulting implementation requires expert knowledge for evaluating implementation trade-offs.
The app development with Outsystems allows for a visual modeling of program actions and control structures as shown in Figure~\ref{fig:uc_app}. The specification of the individual elements as well as other application components required the knowledge of software development concepts, such as variables, datatypes, event listeners, and HTTP request and, in one case, debugging through the logging and interception of requests. On the other hand, Outsystems might be considered a ~\textit{low-code} approach with complex capabilities suitable for developers. 
During development, code consistency, spotting errors visually, discussing and communicating with domain experts, and cross-platform generation have proven beneficial.

\section{Conclusion}

In this paper we reviewed academic model-driven engineering approaches and industrial low-code and no-code platforms for supporting the development of blockchain-based applications. Whereas academic approaches mostly focus on the technical aspects of development, industrial approaches showed a high maturity in terms of usability and integration capabilities. For future developments, more holistic, cloud-based approaches involving business, application, and technical layers seem desirable. With regard to academic approaches, the provision of integration capabilities and sustainable prototypical implementations present current major challenges.

\section*{Acknowledgment}
This work was supported by the Swiss National Science Foundation project Domain-Specific Conceptual Modeling for Distributed Ledger Technologies [196889].


%
%

%
%
%
\bibliographystyle{splncs04}
\bibliography{literature}
\end{document}